\documentclass{amsart}
\usepackage[novbox]{pdfsync}
\usepackage{listings}
\usepackage{color}
\usepackage{pgf,tikz,tcolorbox,chemfig,xcolor,listings}
\usepackage{tikz-3dplot}
\usepackage{tikz-cd}
\usetikzlibrary{shapes,arrows}
\usepackage{url,%authblk,subfigure,
subcaption,
comment,verbatim,float,epstopdf,array,
mathrsfs,xcolor,
amssymb,amsfonts,graphicx,amsmath,multicol,fancybox,import}
\usepackage[version=4]{mhchem}
\usepackage{bm,bbm,stackrel}
\usepackage{etoolbox}
\patchcmd{\MaketitleBox}{\footnotesize\itshape
\elsaddress\par\vskip36pt}{\footnotesize\itshape\elsaddress\par\parbox[b][36pt]
{\linewidth}{\vfill\hfill\textnormal{=day}\hfill\null\vfill}}{}{}%
\patchcmd{\pprintMaketitle}{\footnotesize\itshape\elsaddress\par\vskip36pt}{\footnotesize\itshape\elsaddress\par\parbox[b][36pt]{\linewidth}{\vfill\hfill\textnormal{=day}\hfill\null\vfill}}{}{}%
\usepackage[final]{hyperref}
\usepackage{multirow,lipsum}
\usepackage{array,listings,blkarray,mdframed}
\usepackage{tabularx}
\usepackage{nccmath}
\usepackage[utf8]{inputenc}
\usepackage{todonotes}

\usepackage{graphicx,epsfig}
\graphicspath{{figures/}}
\graphicspath{{Notebook/}}
\usepackage[normalem]{ulem}

\usepackage{lineno}
\usepackage{color}
\usepackage{textcomp}
\RequirePackage{calc}

\usepackage{wasysym}
\usepackage{empheq}
\usepackage{url}
\usepackage{epigraph}
\usepackage{listings}

\usepackage{tikz}
\usetikzlibrary{positioning}
\usetikzlibrary{arrows}
\usetikzlibrary{arrows.meta}
\usetikzlibrary{calc}
\newtheorem{assumption}{Assumption}
\def\beA{\begin{assumption}}\def\eeA{\end{assumption}}

    \makeatletter
\newcommand{\xrightleftarrows}[2]{%
  \mathrel{\mathop{%
    \vcenter{\offinterlineskip\m@th
      \ialign{\hfil##\hfil\cr
        \hphantom{$\scriptstyle\mspace{8mu}{#1}\mspace{8mu}$}\cr
        \rightarrowfill\cr
        \vrule height0pt width 2em\cr
        \leftarrowfill\cr
        \hphantom{$\scriptstyle\mspace{8mu}{#2}\mspace{8mu}$}\cr
        \noalign{\kern-0.3ex}
      }%
    }%
  }\limits^{#1}_{#2}}%
}
\makeatother

\newtheorem{theorem}{Theorem}
\newtheorem{remark}{Remark}
\newtheorem{proposition}{Proposition}
\newtheorem{definition}{Definition}
\newtheorem{corollary}{Corollary}
\newtheorem{example}{Example}
\newtheorem{lemma}{Lemma}

\def\fp{fixed point}\def\mM{{\mathcal M}}\def\mC{{\mathcal C}} \def\mR{{\mathcal R}}
\newtheorem{open}{Open Problem}
\def\eeO{\end{open}} \def\beO{\begin{open}}
 \def\eeD{\end{definition}} \def\beD{\begin{definition}}
\def\beR{\begin{remark}} \def\eeR{\end{remark}}
\def\beL{\begin{lemma}} \def\eeL{\end{lemma}}
\def\beC{\begin{corollary}
  }\def\eeC{\end{corollary}}
  \def\beT{\begin{theorem}}\def\eeT{\end{theorem}}
  \def\beP{\begin{proposition}} \def\eeP{\end{proposition}}
\def\beXa{\begin{example}} \def\eeXa{\end{example}}
\def\beA{\begin{assumption}} \def\eeA{\end{assumption}}

\newtheorem{conjecture}{Conjecture}
\def\beCo{\begin{conjecture}
  }\def\eeCo{\end{conjecture}}
\def\beAs{\begin{ass}
  }
\def\eeAs{\end{ass}}
\newcommand{\R}{\mathbb{R}}
\newcommand{\N}{\mathbb{N}}

\def\How{However, }
\def\bep{\begin{pmatrix}} \def\eep{\end{pmatrix}}
\def\beb{\begin{bmatrix}} \def\eeb{\end{bmatrix}}

\def\NGM{next generation matrix}

\def\l{\lambda}
\def\com{compartment}

\providecommand{\pp}[1]{\left[#1\right]} %[.]
\providecommand{\pr}[1]{\left(#1\right)} %(.)

\definecolor{dkgreen}{rgb}{0,0.6,0}
\definecolor{gray}{rgb}{0.5,0.5,0.5}
\definecolor{mauve}{rgb}{0.58,0,0.82}

\newcommand{\D}{{\color{blue}\bfseries D}}

\def\QED{\hfill {$\square$}\goodbreak \medskip}

\renewcommand{\theta}{\vartheta}

\renewcommand{\thefootnote}{\fnsymbol{footnote}}

\numberwithin{equation}{section}

\def\eig{eigenvalue}\def\satd{satisfied}

\def\bc{\begin{cases}
  }     
\def\ec{\end{cases}}

  \newcommand{\beq}{\begin{eqnarray}
    }

\def\eeq{\end{eqnarray}} 
 
   \newcommand{\be}[1]{\begin{equation}\label{#1}}
\newcommand{\ee}{\end{equation}}
\def\bea{\begin{eqnarray*}}\def\ssec{\subsection}
  \def\prf{{\bf Proof:} } \def\satg{satisfying} 
\def\eea{\end{eqnarray*}} \def\lab{\label} \def\la{\lambda}

\def\mF{\mathcal F} \def\mV{{\mathcal V}}
      
    \def\sats{satisfies}  \def\saty{satisfy}       \def\Oth{On the other hand, } 
\def\ACR{absolute concentration robustness}
 \def\Eq{\Leftrightarrow}

\def\BEN{\begin{enumerate}}  \def\BI{\begin{itemize}}
\def\EEN{\end{enumerate}}   \def\EI{\end{itemize}} \def\im{\item} \def\Lra{\Longrightarrow}  \def\eqr{\eqref}  \def\coe{coefficient}

\def\mR{\mathcal R}

\def\G{\Gamma}%{N}%
\def\g{\gamma}   \def\de{\delta}  \def\b{\beta}

\def\k{\kappa}
\def\Fr{Furthermore, }

   \def\wrt{with respect to }

  \def\resp{respectively}   
\def\mS{{\mathcal S}}  \def\eqr{\eqref}  
\def\wk{well-known}
\def\a{\alpha}%{\; \mathsf a}
\def\l{\; \mathsf l}
\def\fr{\frac} \def\im{\item}

\newcommand{\iem}{\mathsf i}

\def\eeD{\end{definition}} \def\beD{\begin{definition}}
\def\beR{\begin{remark}} \def\eeR{\end{remark}}
\def\beL{\begin{lemma}} \def\eeL{\end{lemma}}

         \def\vi{\vec \iem \;}  
     
     \newcommand{\bff}[1]{{\mbox{\boldmath$#1$}}}
    \def\beP{\begin{proposition}} \def\eeP{\end{proposition}}

    \def\m0{{\mathcal R}_0}  %{I^{(0)}}}
    \def\ME{mathematical epidemiology} \def\bfp{boundary fixed point}

    \long\def\symbolfootnote[#1]#2{
\begingroup
\def\thefootnote{\fnsymbol{footnote}}\footnote[#1]{#2}
\endgroup}

\def\l{\lambda}%\def\e{i_{ref}}
\def\F{F}
 \def\D{\Delta}

\newcommand{\red}{\textcolor[rgb]{1.00,0.00,0.00}}
\newcommand{\blue}{\textcolor[rgb]{0.00,0.00,1.00}}

\def\y{\; \mathsf y} \def\brn{basic reproduction number }
\def\sats{satisfies }\def\beV{\begin{verbatim} }

\def\xy{\mathbf{xy}}\def\x{\mathbf{x}}\def\y{\mathbf{y}}
\def\regS{regular splitting}

\newcommand{\figu}[3]{
\begin{figure}[H]
\centering
\includegraphics[scale=#3]{#1}
\caption{#2\label{f:#1}}
\end{figure}
}
\makeatletter
\renewcommand*\env@matrix[1][*\c@MaxMatrixCols c]{%
  \hskip -\arraycolsep
  \let\@ifnextchar\new@ifnextchar
  \array{#1}}
\makeatother
\def\sm{stoichiometric matrix}  

\def\HJF{Feinberg-Horn-Jackson}
\def\RH{Routh-Hurwitz}
 \def\ch{characteristic polynomial}
\def\cNGM{\cite{Diek,Van,Van08}}
\def\STP{sharp threshold property}

\def\How{However, } 
\def\MH{Metzler-Hurwitz } \def\cNGM{\cite{Diek,Van,Van08}}
\def\CEP{competitive exclusion principle}
\def\GAS{globally asymptotically stable}
\def\LAS{locally asymptotically stable}
\def\EE{endemic point}
\def\brn{basic reproduction number }

 \def\DFE{disease free equilibrium}\def\ie{\i_{ee}} 
\def\MTT{matrix tree theorem}
   \def\Ma{{Mathematica}}

 \def\coo{coordinate}

\def\GAS{globally asymptotically stable}
\def\MM{Michaelis-Menten}
\def\CRN{chemical reaction networks}
 \def\defi{deficiency}

\def\coords{coordinates}

\def\nne{nonnegative}
\def\BIN{biological interaction network }\def\Sm{stoichiometric matrix}\def\Ga{\Gamma}\def\var{variable}\def\admy{admissibility}\def\ie{i.e }\def\jin{{Jacobian matrix of  the infection vector field with respect to the infection variables}}

\def\MAR{mass action representation}\def\rep{representation}
\def\WRZD{weakly reversible, zero deficiency representation}

\vfuzz \hfuzz
\begin{document}

\title[Next Generation Matrix Method for CRNs]{Stability in  Reaction Network  Models via an Extension of the Next Generation Matrix Method
  }
\author{Florin Avram}
\address{Laboratoire de Mathematiques Appliqu\'{e}es, Universit\'{e} de Pau, 64000, Pau, France}
\email{Florin.Avram@univ-Pau.fr}

\author{Rim Adenane}

\address{Department of Mathematics, Faculty of Science,   Ibn-Tofail University, K\'enitra, 14000, Morocco}
\email{rim.adenane@uit.ac.ma}

\author{Andrei D. Halanay}

\address{Department of Mathematics, University of Bucharest, Bucharest, RO-010014, Romania}
\email{halanay@fmi.unibuc.ro}

\author{Matthew D. Johnston}

\address{Department of Mathematics and Computer Science,Lawrence Technological University, Southfield, MI, 48075, USA.}
\email{mjohnsto1@ltu.edu}

 \maketitle

%\date{\parbox{\linewidth}{\centering%=day\endgraf\bigskip }}

%\maketitle
 \begin{abstract}

We investigate structural methods for studying boundary equilibria of chemical reaction network (CRN) models via the Next Generation Matrix (NGM) method of mathematical 
epidemiology, which are based on  symbolic Jacobian decompositions.
%Through the lens of Jacobian splittings and absolute concentration robustness (ACR), we show how the Next Generation Matrix (NGM) method may be generalized to include endemic fixed points.
We apply this to several CRNs known from the literature (including systems with MAK (Mass Action Kinetics) and EnvZ-OmpR models) and explore how simplifications via
network translation yield well-behaved weakly reversible, zero deficiency (WR-ZD) surrogates. Computations are illustrated using our \texttt{Epid-CRN} package in
\textit{Mathematica}. This package provides symbolic tools   for computing bifurcation thresholds and verifying local stability of \fp s  of
positive systems.

 \end{abstract}

\keywords{ Positive  systems; xy models; Chemical Reaction Networks; \ME; Next Generation Matrix method; absolute concentration robustness}

\tableofcontents

\section{Introduction}
The   biological interaction networks (BIN)  sciences, like:
 \BEN \im population dynamics,
 \im ecology,
 \im mathematical epidemiology (ME),     \im virology, \im \CRN\  (CRN),
 etc.
  \EEN
  are mostly concerned with   positive systems, some common themes being multi-stationarity, local and global stability,  bifurcations,  the constructions of Lyapunov functions (LFs), persistence, permanence and extinction, etc.

\beD  We will call positive system
  \cite{rantzer2015scalable} an ODE which leaves invariant the positive orthant.
  \eeD

%\ssec{Motivation}

  However,  due to the different nature of their examples, they have diverged historically, and there exist fundamental mathematical methods and concepts which have impacted one of the sister sciences only. For example:
    \BEN \im
    The {\bf \NGM\ (NGM)}  method \cNGM\  is ubiquitous in ME and  absent in CRN, until the appearance of this paper one year ago, which shows it to be also useful in studying CRN models with ACR (see also the review \cite{AAN}).
    \im Also, the \STP\ (STP), \ie\ the transfer of global asymptotic  stability (GAS) from a fixed boundary point called \DFE\ (DFE) towards a
unique  fixed interior point called \EE\ (EE), has mainly been studied in
ME.
\im \Oth\ the generalization  of  the STP known as \CEP\ (CEP) has originated in population dynamics and   ecology, in the context of the competition of species, but is also studied nowadays in ME \cite{Gavish,Gavish24}, in the context of the competition of virus strains.

\im  CRN theory provides a formal language  which may be applied to all the
  positive systems sciences (this has been obviously known for a very long time), and  it has also  developed remarkable tools, which include the
  {\bf mass action representation},   {\bf complex-balanced  equilibria}, {\bf Horn-Jackson-Feinberg} and {\bf Petri/Volpert/DSR graphs}, the {\bf deficiency},       {\bf robust Lyapunov functions}, etc. The case that it may help in solving ME problems has already been made in \cite{AAV,AAN}.  Below
  we prove the reciprocal: ME methods may be useful for solving CRN models. In fact, we like to see both disciplines as examples of a unifying \BIN\ (BIN) science, which is starting to emerge slowly but surely recently, and whose birth certificate  we like to trace back to \cite{AdLSpetri}.

\EEN

Despite the different points of view mentioned above, we believe that there is
also scope  for unification, by developing a theory of $\xy$ models, first suggested in the pioneering paper \cite{Van}.

Heuristically, positive systems have two types of variables: some which may converge towards $0$, to be called $\x$ variables, and some which may not, to be called $\y$ variables.  The fact that this structural difference may give rise
to computational methods, like the NGM of \cite{Van}, and like the explicit Lyapunov functions of \cite{Fall},  suggested to us  that this structural classification might yield results outside ME (where it originated). Indeed, this paper shows that NGM yields results which may not be accessible otherwise,  for CRN models exhibiting \ACR\ (ACR).

Before delving into that, we open a parenthesis.  The results described below were obtained in parallel with the creation of the Mathematica package EpidCRN, which required first making a choice on how to input models.

\subsection{How to represent CRN and  ME  ODE models via symbolic software?}
{For parameterizing ME and CRN ODE models},
   there are three natural choices:

 \BEN \im
 The parametrization   used traditionally for ODE models   is as a pair (RHS, var), where RHS denotes the set of formulas for the derivatives, and var are the  variables whose rates of change we study.
 This parametrization is sufficient for solving small size symbolic bifurcation problems. It cannot go beyond that however, since it ignores the ``network" (or graph) structure revealed by the following two parametrizations.

  \im Applying chemical reaction networks methods  requires that a model be defined as a triple $(\G, {\bf R}, x)$, so that \be{crnsys}x'=RHS=\G .{\bf R}(x)\ee
  \BEN \im where $x$ denotes the variables \im
   $\G$ is the ``stoichiometric matrix'' (SM), whose columns  represent directions in which several species/\com s change simultaneously,
   \im  ${\bf R}(x)$ is the  vector of rates of change associated to each direction, also known as kinetics, assumed all to be \nne.
    \EEN

For example, the  simplest
 SIRS ODE without demography is defined by the  triple $(\G, {\bf R}, x)$:
  \be{SIRc}
\dot x=\bep
\dot s \\ \dot i    \\ \dot r\eep
 =  \bep - 1&0&1&-1 \\
  1&- 1&0&0\\
  0&1&- 1&1
\eep \bep
\b s  i    \\ \g_i i    \\ \g_r r \\ \g_s s  \eep:=\Gamma {\bf R}(x).
\ee

Note this representation is also the first step towards defining an associated CTMC model on the integers.

This parametrization  has had remarkable successes, including recently  for studying ``robust questions" like  the existence of ``robust Lyapunov functions " which depend only on the ``reaction variables $\bf z={\bf R}(x)$~\cite{al2014robust,BG14,BlaGio17poly,Franco,AAScomp}. All these papers
assume that the rates $R(x)$ certain admissibility conditions --see  Ref. \cite{ARS24} and  definition \ref{admG}.
\iffalse
\beD \lab{admG}
 The $\Ga$ \admy\ conditions for a CRN defined by its \Sm\  are:\begin{enumerate}
 		\item[\bf {A1.}] $R_j(x)$ is continuously differentiable, $j=1,..,n_r$, in an open set containing the \nne\ orthant;
 		\item[\bf {A2.}]  $ \bc \Gamma_{ij}< 0\\x_i=0 \ec \Lra R_j(x)=0$; (this is necessary for positivity, by the "Hungarian lemma" \cite{hun}
 		\item[\bf {A3.}] $\bc \Gamma_{ij}< 0 \Lra {\partial R_j}/{\partial x_i}(x) \ge 0\\\Gamma_{ij}\ge 0 \Lra {\partial R_j}/{\partial x_i}(x)\equiv 0\ec $;
 	
 		\item[\bf {A4.}] The first inequality in   {A3} holds strictly for all positive concentrations, i.e., when $x \in \mathbb R_+^n$.
 	 \EEN

\eeD

\beR Unfortunately, the ODE \eqr{SIRc} does not belong to the class of ``$\Ga$ admissible" models.  It does not \saty\ the second $A3$ \admy\ condition above,
 since $\b s i$ depends on $i$. In fact, A3 is not  \satd\  for any ME model we know of. \eeR

   Applying CRN results to  ME models may require a richer \rep\ of CRNs, provided in the next item.

%%%%%%%%%%%%%%%%%%%%%
\fi
   \im The third parametrization, the traditional one used in CRNT,   replaces each column of $\G$ by a ``reaction". For example, the second column in \eqr{SIRc} is replaced 
 by $I->R$. The first is replaced by $S+ I->2 I$ (not by $S-> I$), due to the mass-action representation, recalled later. This the most human-error proof method of entering large 
systems with tens of reactions. The end result is replacing  $\G$ by the difference of two \nne\ input and output matrices, such that  \be{ab}\G=\beta-\alpha.\ee

\EEN
   \beR  Note  that the representation \eqr{ab} is  the first step towards defining an associated DTMC (discrete time Markov chain) model on the integers, where we distinguish between in and outflows in a state, which is only possible in discrete time.

   Note also that to one stoichiometric matrix  $\G$  one may associate an infinite family of $(\alpha,\beta)$ models which \saty\ $\beta-\alpha=\Ga$, which have  different mass-action rates.
  \eeR

   In the third \rep,   each column of $\G$/``reaction" is
   associated   to a  directed pair consisting of a ``source complex'',  and a ``product complex", which we proceed now to    define.

    \beD \lab{d:CRN} A %{\bf mass-action}
    CRN is defined by a triple $\{\mathcal{S},\mathcal{C},\mathcal{R}\}$, where $\mathcal{S},\mathcal{C},\mathcal{R}$ are the set of species,
    complexes and reactions respectively.

\begin{align*}
    \mathcal{S} &= \{S_1, ... , S_i,..., S_{|\mathcal{S}|}\}\\
    \mathcal{C} & = \{y_1,...,y_\alpha,...,y_{|\mathcal{C}|} :y_\alpha \in \mathbb{N}^{|\mathcal{S}|}\}\\
    \mathcal{R} &= \{ \ce{ y_{\alpha} ->[k_{y_{\alpha} \to y_{\beta}} * x^{y_{\alpha}}] y_{\beta} } : k_{y_{\alpha} \to y_{\beta}} \geq 0\},
\end{align*}
where Roman letters ($i,j$) and Greek letters ($\alpha,\beta$) are used to denote species and complex indices, respectively, and $x=\{s_1, ... , s_i,..., s_{|\mathcal{S}|}\} $ denotes the vector of unknowns.
A complex is a multi-set of species, and is denoted by the column vector $y_\alpha$ representing the stoichiometry/direction of the multi-set.
\eeD
  \beXa {The SIRS ODE without demography \eqr{SIRc} is induced by the reactions:}
{
\be{RNSIR} \bc  \ce{ S + I ->[\beta s i] 2I}\\ \ce{ I ->[\gamma_i   i] R } \\
 \ce{R ->[\g_{r} r] S } \\
 \ce{  S  ->[\g_{s} s] R } \ec \ee
}

Here, \begin{align*}
    \mathcal{S} &= \{S,I,R\}, x=(s,i,r),\\
    \mathcal{C} & = \{y_1=(1,1,0),y_2=(0,2,0),y_3=(1,0,0),y_4=(0,1,0),y_5=(0,0,1) \}\\
    \mathcal{R} &= \{ \ce{ y_1 ->[\b  x^{y_1}=\b s i] y_2}, \ce{ y_4 ->[\g_i  i] y_5},...\},
\end{align*}
\eeXa

\beR The complexes of the last three reactions of \eqr{RNSIR} are self-explanatory, but not those of the first (see formulas of $y_1,y_2$).  The problem is that the natural ``flow diagram" \rep\ $S\rightarrow I$ is $\Gamma$ unadmissible.
To overcome this, we may introduce artificially any positive multiple of $I$ in the first reaction, on both sides; it is easily checked that all these models  will enjoy the   ``extended admissibility" conditions AK below.  The choice we made also
ensures that our SIRS is a \MAR, \ie\ that the linear coefficients multiplying $S,I$ on the LHS of  \eqr{RNSIR} equal the exponents of $s,i$ in the rate of the reaction.
\eeR

The following conditions are modeled after \cite{AdLSpetri}.

\beD \lab{admG}
The  \admy\ conditions for a CRN defined by its input and output matrices $\a, \b$ are:

\begin{enumerate}
		\item[\!\!] {\bf AK1}. each reaction varies smoothly with respects to its reactants, i.e $R(x)$ is $\mathscr C^1$;
		\item[\!\!] {\bf AK2}. a reaction requires all its reactants to occur, i.e.,  if $\alpha_{ij}>0$, then $x_i=0$ implies  $R_j(x)=0$; %
		\item[\!\!] {\bf AK3}.  a reaction rate increases if a reactant increases: \begin{gather*}\frac{\partial R_i(x)}{\partial x_j}
    \begin{cases}
        \geq  0 &  \text{ if $x_j$ is a reactant, \ie } \alpha_{ij}>0\\
%        \leq 0 &\text{ if $x_j$ is a product,\ie } \beta_{ij}>0  \\
        = 0  &\text{otherwise}
    \end{cases}
    \end{gather*}
    Furthermore, the aforementioned inequality is strict whenever the reactants are strictly positive:
    \begin{gather*}
    \frac{\partial R_i(x)}{\partial x_j}  >0 \text{ if $x_j$ is a reactant and all the reactants are present}.% \\
%        <0 & \text{if $x_j$ is a product and all the products are present}
%    \end{cases}
  \end{gather*}
		
	\end{enumerate}

\eeD

From the extended representation,  one may construct $\a,\b$, and the stoichiometric/directions matrix $\Gamma=\b-\a$, by associating to each reaction the column   vectors given by its right and left complexes $y_{\beta}, y_{\alpha}$, and  finally the column   vector  $ y_{\beta}-y_{\alpha}$.

As for the implementation solution we adopted in Epid-CRN,   it is to input only $\G$ for models which
do not exhibit direct autocatalysis (\ie the presence of the same species on both sides of a reaction), while models which exhibit direct autocatalysis we input the pair $(\a,\b)$; we follow here the  Matlab package {LEARN} \cite{AAnew,AAScomp,malirdwi_LEARN,
Ali23gr}, available at \url{https://github.com/malirdwi/LEARN}.
\iffalse
Remarkable tools and theorems concerning intriguing phenomena have also been discovered in ME and in ecology:
  \BEN \im
    The {\bf \NGM\ (NGM)} reduction method \cNGM for the local asymptotic  stability (LAS) of boundary points, which ends up being  expressed as
  \begin{equation}\label{R0fr}
  R_0 <1
\end{equation}in terms of the "\brn\ $R_0$ (via {\bf regular splittings}~\cite{Plemmons,Fall,thieme2009spectral}).

Note this result   is ubiquitous in ME and almost absent outside this field (see though \cite{AAN,AAHJ}, where it was shown to be also useful in studying CRN models with ACR).
    \im Also, the STP, the transfer of global asymptotic  stability (GAS) from a fixed boundary point called \DFE\ (DFE) towards a
unique  fixed interior point called \EE\ (EE), has mainly been studied in
ME.
\im \Oth\ the generalization  of  the STP known as \CEP\ (CEP) has originated in population dynamics and   ecology, in the context of the competition of species, but is also studied nowadays in ME \cite{Gavish,Gavish24}, in the context of the competition of virus strains.
\EEN

  Given all the remarkable results concerning stability in CRN and ME,
  \fi

\ssec{Fundamentals  of $xy$ models}
 We will be concerned below with boundary fixed points of positive systems. They belong  typically to invariant boundary faces which are non-trivial in the sense that their non-zero \coo  s do not contain a conservation. %The \corr\ non-zero \coo  s are called a ``critical siphon" in \CRN\ theory.

Assume a positive system with species $ \mathcal{S}$ admits boundary fixed points,  and let  $\mathcal{X} \subset \mathcal{S}$ denote the  set of zero \coo  s for one boundary fixed point, assumed to be maximal in the sense
$card(\mathcal{X})\geq card(\mathcal{X}')$, for any $\mathcal{X}' $ which is the  set of zero \coo  s for any other boundary fixed point. We reorder the species set so that the first $m$ species are elements of $\mathcal{X}$ (i.e. $X_i \in \mathcal{X}$ for $i=1, \ldots, m$) and the remaining $n-m$ are elements of its complement $\mathcal{Y}$. Let $\mathbf{x} \in \mathbb{R}^m_{\geq 0}$ denote the \coo  s of species in $\mathcal{X}$ and $\mathbf{y} \in \mathbb{R}^{n-m}_{\geq 0}$ denote the \coo  s of species in   $\mathcal{Y}$.

In particular,  it holds that
\beL
\label{lemma1}
The invariance condition for a boundary face of a \textbf{positive ODE system} \[
\frac{dx}{dt} = f(x), x \in \mathbb{R}^n_{>0},
\]
where $ f(x) $ is a smooth vector field, can be expressed as follows.

Let $ \mathcal{X} $ be a boundary face defined by setting some coordinates to zero, e.g.,
\[
\mathcal{X} = \{ x \in \mathbb{R}^n_{\geq 0} \mid x_i = 0 \text{ for } i \in I \}
\]
for some index set $ I \subseteq \{1, \dots, n\} $.
The face $ \mathcal{X} $ is \textbf{invariant} if, whenever a trajectory starts in $ \mathcal{X} $, it remains in $ \mathcal{X} $ for all future times, \ie:

\be{inv}
f_i(x) = 0 \quad \text{for all } x \in \mathcal{X} \text{ and for all } i \in I \Eq f(x) \cdot \nu_X(x) = 0, \quad \forall x \in \mathcal{X},
\ee
where $ \nu_X(x) $ is the outward normal to the face $ \mathcal{X}$ at $ x $.
\eeL

 Let now $\mathbf{x},\mathbf{y}$ denote the \var s which define a maximal invariant facet, and its complement, \resp. We rewrite our system of differential equations as
\begin{equation}
    \label{de2}
    \left\{ \; \; \;
    \begin{aligned}
    \frac{d\mathbf{x}}{dt} & = \mathbf{f}(\mathbf{x},\mathbf{y})= M(\mathbf{x},\mathbf{y}). \mathbf{x}\\
    \frac{d\mathbf{y}}{dt} & = \mathbf{g}(\mathbf{x},\mathbf{y}).
    \end{aligned}
    \right.
\end{equation}
\beD A system which may be written in the form \eqref{de2}, \wrt\ some maximal set $\mathcal{X}$,  will be called below an $xy$ model.
\eeD
The decomposition \eqref{de2} (analogous to the decomposition in \cite{Van} where the species in $\mathbf{x}$ are interpreted as infected/diseased and the species in $\mathbf{y}$ are susceptibles/substrate/noninfected) implies
 the existence of a steady state of the form $(\mathbf{0},{\mathbf{y}}_0) \in \mathbb{R}^n$, and suggests the possible existence of other
 steady states \satg
 \be{dM} det(M(\mathbf{x},\mathbf{y}))=0.\ee

 \beR  The matrix $M(\mathbf{x},\mathbf{y})$ is not uniquely defined. In quadratic models, which are our main object of concern, it coincides with the \jin\ $J_x(\mathbf{x},\mathbf{y})$. \eeR

As a preview of what follows, we end this section with a question:
%\beO What is the mathematical definition of the most convenient disease set for applying NGM?  And how about CRN models, in the absence of exterior inputs, and in the presence of  conservations (which determine the non zero components of the DFE)?\eeO

\beO Does the NGM theory apply to all boundary \fp s? If not,  to which subclass of them?
\eeO
%We will also make the case   that putting together these elements of  three disciplines  leads to an elegant theory in which the sum is bigger than its parts.

\ssec{Regular splitting theorem of \cite{Varga}}

\beD  The spectral radius of a matrix A  is defined by
$$\rho(A) = max\{|\la| ,  \la \in  Sp(A)\},$$
where  $Sp(A)$ denotes the spectrum of A.

The spectral bound of a matrix $A$ is
\begin{equation*}
  s(A) = \max \{ \Re \lambda | \lambda \in Sp(A)\}
\end{equation*}
\eeD

\beD A matrix $A$ is called nonnegative (positive) if $A(i,j) \geq 0 (>0)$ for all $i,j$. This relation will be denoted by $A \geq (>) O.$\eeD

\beD A matrix $A$ is called Hurwitz stable if $s(A) < 0.$ \eeD

\beD A Metzler matrix A is a matrix such that $i \neq j \Lra A(i, j) \ge 0.$  These
matrices are also called quasi-positive matrices. A matrix $A$ such that $-A$ is Metzler is called an M-matrix.
\eeD

\beA The matrix $M=J_x$ appearing in the ``siphon equations" \eqr{de2} for $\bff{x}'$ must be  Metzler, for any $\mathbf{x} \in \mathbb{R}_{\geq 0}^{m}, \mathbf{y} \in
\mathbb{R}_{\geq 0}^{n-m}$.
\eeA

\beR Note that saying $M$ is  Metzler is equivalent to saying that the flow of the siphon \var s, when the other \var s are fixed,  is monotone.
\eeR

Recall that
\begin{lemma}\cite[Lemma 6.12]{diekmann2000mathematical}\label{ld}
  A Metzler matrix $M$ is Hurwitz stable if and only if   $-M^{-1} \geq 0.$
\end{lemma}
\iffalse
\red{Example: $M=\bep -\la_1,0\\0,-\la_2\eep, \la_i >0, i=1,2$ is  Metzler.
Note $\det(M) > 0$.}
\fi

\beD\cite[Def. 3.28]{Varga}
Let $M$ be a square, real Metzler matrix. A regular splitting of  $M$ is a decomposition $M=F+A$, with $\red{-}A^{-1} \geq 0$ and $F \geq 0$ (such a decomposition need not exist, nor be unique).
\eeD

% \beD  A  regular
% splitting is a decomposition of  a real Metzler matrix M   of the form  $$M = A + F,$$ where $A$ is a \MH\  matrix and $F \geq 0$ is a nonnegative matrix.
% (Such \dec\ needs neither exist, nor be unique). \eeD

The major result about regular splittings is
a generalization of the following \wk\ result:
\beL
 Let $J=F-s I_d$ be a real square matrix.
 Then, $J$ is stable  iff
 $s> \rho ( F) $.
\eeL

Varga's result extends the above to the case when $-s I_d$ is replaced by an arbitrary \MH\ matrix.
\beP \cite[Thm 3.29]{Varga},\cite{thieme2009spectral} \lab{p:Var}
 Let $M$ be a real Metzler matrix  admitting regular splitting(s).
 Then $M$ is \MH\ (or, equivalently, $-M^{-1} \geq 0$)  if and only if
 \begin{equation}\label{reg-split}
  \rho\left( F (-A)^{- 1}\right)=\frac{\rho\left(-M^{-1}F\right)}
  {1+\rho\left(-M^{-1}F\right)} <1,
\end{equation}
  where $M = A + F$ is  any regular splitting of  $M $.
\eeP

\prf Because $M$ has a regular splitting, then
$$M  = \pp{F (-A)^{-1} -I_d}(-A), $$
with  $-A^{-1} \geq 0, F (-A)^{-1}\geq 0$.

Now $\rho(-FA^{-1}) < 1, F (-A)^{-1}\geq 0$ imply that $ I_d + F A^{-1}$ is invertible and
$\left( I_d + F A^{-1}\right)^{-1} \geq 0$ \cite[Thm 3.16]{Varga} (this is a generalization of the scalar geometric series theorem). \Fr $$M^{-1}=A^{-1}
\left( I_d + FA^{-1}\right)^{-1}, $$
and being the product of two positive matrices, $M^{-1}$ is positive.

On the other hand if $M^{-1} \geq 0$ and $M=A+F$ is a regular splitting, let $G=M^{-1}F'$. Then $A^{-1}F'\geq O,$ $G \geq O$ and both matrices have common
eigenvectors and the eigenvalues $\mu$ of $A^{-1}F'$ and $\tau$ of $G$ corresponding to the same eigenvector satisfy
\begin{equation*}
  \mu = \frac{\tau}{1+\tau}.
\end{equation*}
Both $A^{-1}F'$ and $G$ are nonnegative so using Perron-Frobenius theorem one can consider only those eigenvectors $x$ which are nonnegative such that $\tau \geq 0.$ Then $\mu$
is a strictly increasing function of $\tau$ and is maximized by $\tau = \rho(G).$ This gives equation \ref{reg-split}.
\QED

% \beC  \lab{c:NGM} The DFE is locally asymptotically stable (LAS) if $$R_0(F, A):=\rho(F (-A)^{- 1}) <1,$$ where $J_x=F+A$ is any regular splitting of the \jin\ $J_x$. \eeC
% \prf Follows from Proposition \ref{p:Var}. \QED

\ssec{Can CRN and ME methods  be useful also outside their field?}

 The answer to this natural  question was  not at all obvious until recently.
 While the folklore knowledge
  that such a transfer of information should be  possible may be traced back
 to at least 50 years ago,  the moment when this transfer actually started, is quite recent -- see  \cite{giordano2020sidarthe,blanchini2021generalized,BG21} and previous papers of Giordano, Blanchini and Colaneri.

  We have also provided some interactions,  in the
 recent paper Vassena \& al \cite{AAV}: \BEN \im One remarkable recent  result of Refs. \cite{Vas,VasStad} states that {\bf any auto-catalytic CRN will admit  unstable equilibria (increasing thus the chance of Hopf bifurcations), if all the reaction rates are ``rich enough"}, for example \MM\ (MM). This was applied in \cite[Thm 3.1]{AAV} to ME.  This result  may be informally stated as:
      Consider any epidemiological system in which there exists an ``$S\to I$ infection  reaction" with ``admissible rate" $R_1(s,i)$, and an ``$I\to ...$ treatment  reaction" with
      admissible rate $R_2(i)$, where admissible is defined in the original papers.
      %, and also here below, in Remark \ref{r:adm}.
      Then, the ``symbolic Jacobian'', in which $R_1,R_2$ are not specified (but the sign of their derivatives
      is specified, via admissibility), may always have purely imaginary eigenvalues if $R_1,R_2$ are ``rich'' in the sense of Refs. \cite{Vas,VasStad} (for example, \MM).

      Briefly,
 all  epidemic models admit ``symbolic  bifurcations", provided their rate functions  have enough parameters;  it is only the restriction
to mass action that may prevent the occurrence of Hopf bifurcations.

Note that this result had been observed  empirically in many particular three compartment SIR-type models, but the fact that the  number of  compartments and the exact architecture of the model are irrelevant, was not understood.

     Note also that since in ME it does  not make  much sense to replace  linear rates by \MM,  the result  in Ref. \cite{AAV}, which only assumes the ``relevant'' rates to be rich,
        required a careful  reexamination of the method of Refs. \cite{Vas,VasStad}.
       % A related example is provided here in section \ref{s:SIRScap}.
       \im A second CRN result  exploited in Ref. \cite{AAV} was the {\bf Inheritance of oscillation in chemical reaction networks} of Ref. \cite{banaji2018inheritance},
       which gives conditions for Hopf  bifurcations to be inherited by models, given that they exist in a sub-model where some parameters are $0$.
  This allowed establishing the existence of  Hopf  bifurcations for the mass-action ME model %\eqr{SIRn} below
  (using their existence in the case of no demography, already established by Hethcote and Van den Driesche \cite{HethSVDD}).
  \im In the opposite direction, we show below, (see also the review paper \cite{AAN}, which was  submitted simultaneously with this one), that the \NGM\ method of
  Refs. \cite{Diek,Van,Van08}, may be  applied to find conditions for  the stability
  of fixed boundary points of certain CRN models.
\EEN

We end the introduction by noting  that a definition of ME models is still lacking.
%One might want to ask for essential non-negativity, but  for a counterexample see Ref. \cite{Rodrigues}.
We might  want to rule out types of reactions absent in all
epidemiological models, like simultaneous degradations (deaths), written by chemists in the form ``A''+``B''$\to$0, and ``two from one reactions'', like ``C''$\to$``A''+``B''. It seems premature now   to  speculate on this question;
 however, as a wide digression, we conjecture  that
\beO \lab{o:MEm}
A definition of mass-action ME models should identify an appropriate  subclass of the class of
chemical reaction networks, possibly the subclass including only transfers (monomolecular reactions), and bimolecular reactions of the types $S+E\to 2 E, S+I\to I+E$, as encountered in SEIR.\eeO
%systems admitting a BDC-decomposition in the sense of \cite{GiuliaFranco}, which includes already all , and also flow and compartmental systems (for a survey see \cite{Giulia}).  We identify this as a direction for future work.

%%%%%%%%%%%%%%%%%%%%%%%%%%%%%%%%%
\ssec{Contents} Since this essay %, in its complete ArXiv version,
is in part a notebook accompanied manual
for solving ME and CRN DFE stability problems (see the notebooks associated to each section, available at \url{https://github.com/florinav/EpidCRNmodels }), one  issue  we encountered was ordering the problems by their complexity.
For that, we found useful  the CRN classification of complexity in Ref. \cite{BaBo23}, according to $(n_{Species}, n_{Reactions},rank(\Gamma))$, but we added at the end the deficiency.
\beD  The deficiency of a CRN is defined to be the number of complexes $|\mathcal{C}|$ minus the rank of the \sm, and minus the number of components in the \HJF\ graph (for further
information on the deficiency theorems and CRN's in general, see for example the tutorials Refs. \cite{Gun,Ang,Yu,Cox}).
\eeD

%We  offer in the associated GitHub repository \url{https://github.com/florinav/EpidCRNmodels }  a Mathematica package, Epid-CRN,  which
%  is addressed to researchers of both disciplines, and  provide illustrative notebooks, which in particular solve  a few minor open ME and CRN problems.
%Our package may also be used to study  easy cases of  analogue continuous time Markov chain (CTMC) ME and CRN  models, as explained
%in  the Arxiv version of the paper, called Investigating synergies  between
%  chemical reaction networks (CRN)
%  and  mathematical epidemiology (ME), using the  Mathematica package Epid-CRN.

\iffalse
 We  start in section \ref{s:SIR} with  some simple versions of the SIR model, included with the purpose of disseminating the basics of ME.
Section \ref{s:SIRS} provides a  notebook for a mass-action three species, eight reactions SIRS model, where the application of a Routh Hurwitz -type criterion for the unique positive endemic point is straightforward, and rules out Hopf bifurcations.
The notebook for section \ref{s:SIRScap} offers an example  with Hopf bifurcation
 for a SIRS model with one functional rate $f(i)$, in terms of the values $f(i), f'(i)$ at the endemic point. The example may be realized by choosing a Michaelis-Menten  type $f(i)$.
\fi

Section \ref{s:NGM} complements the NGM method \cNGM\ by a NGM heuristic, which works sometimes when the former can't be applied.
Section \ref{s:ACR} provides a quick introduction to CRN models displaying absolute concentration robustness (ACR)-- see Definition \ref{d:ACR}.
 Section \ref{s:ACRex}  tackles the stability  of boundary points of several CRN models with ACR, via the NGM method and heuristic.
  %{s:SIRSrob} lists results obtained with a Matlab LEARN notebook, for a robust SIRS model with no demography.

Section \ref{s:Ton} discusses in parallel a  CRN example, and one of its \WRZD\ realizations. This is  just an example of one of  several network reduction techniques which turned out useful in CRN; our aim is to inform ME researchers
about this potentially interesting set of methods.

\section{The \NGM\ method for determining the   stability
 threshold of the DFE}\lab{s:NGM}
 We discuss here the    representation
\be{R0fr} R_0<1\ee
for the stability domain of the Disease free equilibrium (DFE) (the \fp\  having a maximal number of variables which equal $0$), via the fascinating  \NGM\ (NGM) method \cite{Diek,Van,Van08} (called this way since it replaces the investigation of the Jacobian by that of a matrix whose origins lie in  the theory of branching processes).

 We summarize first some standard notations used in the literature:
 \BI
 \im DFE = disease-free equilibrium, having a maximal number of 0 coordinates (``infection \var s");
\im EE = endemic, strictly positive equilibrium;
\im equilibria differerent from DFE and EE will be called boundary endemic points;
\im M is matrix appearing in the factorization \eqr{de2}
\im $J_x$ (or $J_I$) is the  Jacobian block corresponding to the x \var s;
\im  F = positive non-constant terms in $J_x$  (``new infection);
\im  A = -V= all the terms in M which are not in F.
\EI

Once the DFE is identified, its  stability
 threshold   may be determined via the NGM method,
 which has two steps:
 \BEN \im The first  is the projection of the Jacobian on the subset of infection equations variables,
  \ie\ the equations involving the $0$ coordinates of the DFE.
  This  may be justified by \BEN
 \im verifying the conditions of \cNGM, or, in particular cases, by\im
 factoring the \ch\ of the Jacobian matrix,  and by eliminating trivial factors (which correspond typically to the non-disease \var s).

 \EEN

 \im  The second is applying \regS\ to the
projected Jacobian $J_x$. Assuming the \jin\ $J_x$ is a Metzler matrix,
 and that it admits a regular splitting $J_x=F +A$, the  study of  $J_x$ is replaced by that of the "\NGM"
 $$ K= F  (-A)^{-1}$$
  which may  be obtained by the algebraic trick:
 $$J_x=F +A=\pp{F  (-A)^{-1}-I} (-A).$$

   The final result is the stability threshold
   \be{R0fr}R_0:=\rho(K)<1\ee\lab{c:NGM}
 (a corollary of Varga's regular splitting Lemma \cite[Thm 3.28]{Varga} -- see also \cite{Berman,Fall, thieme2009spectral}).

 We will call this equation the first law of \ME.
 \EEN

\beR \BEN \im One weak point of the NGM method is the lack of a mathematical definition and discussion of the uniqueness
of the DFE (which is informally,  a fixed point where all  infection variables are 0). Its creators in fact argued sometimes that this question, together with the determination of  ``new infections" interactions, must be resolved in collaboration with an expert epidemiologist.  We  complement this by the following definitions.

\im Another weak point
is that the NGM conditions of \cite{Diek,Van,Van08} are not necessary, as demonstrated by the ACR example of Section \ref{s:ShiFei}, where the condition
\cite[(A 4)]{Van08} is not \satd. This seems  to be the first such example in the literature.
\EEN \eeR

\beD A boundary fixed point whose support (set of zero/infection variables) includes the supports of all other \bfp s will be called DFE.\eeD
\beD \lab{d:ni} The ``new infections" matrix $F$ is the matrix which retains from the \jin\ all the nonconstant positive terms, and is 0 elsewhere.\eeD

This definition is implemented in EpidCRN, and was shown in \cite{AABJ} to produce the ``epidemiologic new infections and $R_0$"~\cNGM\ in  a wide variety of ME examples.

We complement now the  ``epidemiologic new infections" $F$ in definition \ref{d:ni} by an alternative \regS.

\beD A   matrix $F$ of    positive  terms
which provides a regular splitting of $J_x$, i.e. is such that $V=F-J_x$ has only \nne\ off diagonal terms, and such that $V^{-1}$ has only positive terms
(these are related to ``expected dwell times"  of associated CTMC (continuous time Markov chain) models \cite{Hurtado,AAB}),
 will be called {\bf optimal} when $F$ has minimal rank among all the matrices with the above properties.\eeD
\beT The existence of an optimal \regS\ $(F,V)$ of rank one implies that the stability
domain of the DFE is of codimension 1, of the form $R_0(F) <1$. If furthermore the DFE is rational, so is  $R_0(F)$.
\eeT
We will search for an optimal \regS\ of rank one whenever the $R_0$ produced by the  definition \ref{d:ni} is non-rational.

\ssec{A NGM heuristic method}\lab{s:heu}

We offer here  a  ``NGM heuristic" for the projection step, which may hold under wider conditions than those of Refs. \cite{Diek,Van,Van08}, remaining to be investigated.

Our NGM heuristic consists of the following steps:
\BEN \im Write the infection equations  in the form
\be{M}\vi'= M \vi, \ee
where $M$ is a Metzler matrix.
\im Find an optimal \regS\  $M=F-V$,
 to replace the study of the spectrum of Jacobian $M$ ( \wrt\
the infection variables)
by that of the ``next generation matrix" defined by
\be{K} K=FV^{-1}.\ee
\im Determine the spectral radius of $K$, the so called \brn\ $R_0$.\EEN

\beR The first step is suggested by the well-known fact that infection variables are fast compared  to the others, so the non-infection variables may be taken as fixed and their equations ignored, asymptotically.

In connection to the third step, we mention  another empirical observation, that the \ch\ of the matrix $K$ has typically more $0$ eigenvalues than the \ch\ of the Jacobian of the infection equations, which explains the popularity of NGM.

In fact, the overwhelming majority of ME papers study models where all
the eigenvalues of $K$ but one  are $0$, \ie\ K has rank one. This entails a rational formula
and an ``epidemiologic/probabilistic interpretation" for $R_0$.  In this case a simple $R_0$ formula due to
Ref. \cite{Arino} is available (see also Ref. \cite{AABBGH}); however, the observation that this formula is useful for producing rational $R_0$'s even when the  epidemiologic $R_0$ isn't rational seems new.
\iffalse
Two other  cases where $R_0$ has an  interpretation are those where the \ch\ of $K$ fully factors into linear terms (this is obtained in multi-strain models), and those where after removing the $0$ eigenvalues, the \ch\ is a quadratic symmetric polynomial. The latter case,
where $R_0$ is a square root, is encountered below in Section \ref{s:Ton}.

Let us mention also that in Section \ref{s:EnvJ}  we encounter ``a non-interpretable $R_0$", which \sats\ an irreducible cubic equation.
\fi
\eeR

The ``NGM heuristic" has also a ``shortcut rank-one NGM method", which consists of using a rank one regular splitting matrix. These obtain sometimes the correct stability domain of the DFE, raising an  open question (among many others):
  \beO Provide  conditions for  the NGM heuristic and   rank-one optimal NGM methods  to obtain the correct stability domain of the DFE.
  \eeO

\beR The considerations above are explored concretely in the IDHKP, EnvZ-OmpR, and MAK examples that follow.\eeR

\section{A bird's eye view on ACR \cite{ShiFei,karp2012complex,DexterGun,AliciaInv,tonello2018network,
PuenteJohn,joshi2024bifunctional}}\lab{s:ACR}
{A well-studied class of CRNs which frequently have boundary steady states are those with absolute concentration robustness.}
\beD
\label{d:ACR}
A mass action system $\mM = (\mS, \mC, \mR, k)$ is said to have:
\begin{enumerate}
\item
a \emph{robust ratio} between complexes $y$ and $y'$ if the ratio $\mathbf{x}^{y}/\mathbf{x}^{y'}$ takes the same value at every positive steady state $\mathbf{x} \in \mathbb{R}_{> 0 }^n$.
\item
\emph{absolute concentration robustness} in species $A$ if $x_A$ takes the same value at every positive steady state $\mathbf{x} \in \mathbb{R}_{> 0 }^n$. The value $x_A$ is called the \textit{ACR value} of $X_A$.
\end{enumerate}
\eeD
\noindent Note  that $x_A$ is independent of   the conservation constants and of the initial conditions, but this value does depend on the reaction constants $k$, and  that if two complexes $y$ and $y'$ have a robust ratio and differ only in a single species $A$, then ACR is guaranteed for $x_A$ \cite{ShiFei}.

ACR was first rigorously studied by Shinar and Feinberg~\cite{ShiFei} (interestingly, the simplest example they discuss is the  ME model with no demography known as SI, which they
call ``toy model''). Recently, their sufficient conditions have been extended in Ref. ~\cite{joshi2024bifunctional}.

 \beXa The  \ME\ model SI, called ``toy model'' in Ref. \cite{ShiFei}, is:
$$\{
S+I \xrightarrow{k_1} 2I, ~ I \xrightarrow{k_2}S\}.$$  This system has exactly one positive steady state $(k_2/k_1, N_0 - k_2/k_1)$, where $N_0$ is the initial total population, and hence the network has  ACR in the first species $S$.
\eeXa

%Another  definition involving {\bf dynamic  ACR} was offered in~\cite{joshi-craciun,joshi-craciun-2}.

The first sufficient conditions for ACR, due to Shinar and Feinberg \cite{ShiFei}, have been restated in Ref. \cite{tonello2018network} as follows:
\beP
\label{thm:robustness_def0_def1}
Consider a  mass action system $\mM$.
\begin{enumerate}
\item
  If $\mM$ has a deficiency of zero and is weakly reversible,
  then it  has a robust ratio in each pair of complexes $y$ and $y'$ belonging to a common linkage class.
\item
  If $\mM$ has a deficiency of one and admits a positive steady state,
  then $\mM$ has a robust ratio in every pair of nonterminal complexes $y$ and $y'$ in $\N$.
\end{enumerate}
\eeP

ACR has also been analyzed using methods from computational
algebraic geometry~\cite{kaihnsa2024absolute, karp2012complex, beatriz,PuenteJohn}
(see also Refs. ~\cite{case-study,CraStu}
for related works).
This is natural, in view of the fact that
 the ACR property is related to the ideal of the positive steady state locus~\cite{karp2012complex}.

\begin{proposition}[ACR and ideals] \label{prop:ideal}
A mass-action system $( \mM, \kappa)$ has ACR in species $X_i$
if and only if
$x_i - \alpha$, for some $\alpha >0$, is in the ideal of the positive steady state locus of $(\mM,\kappa)$.
\end{proposition}

  Conservative ACR systems  have always boundary steady states, owing to the following result.

\beT If a biochemical system admits: 1) an ACR species $Y,$ with ACR value $y^*$, and 2) a positive conservation law,  then  it must have a unique
fixed point $x^*$   on the boundary of the positive orthant, of minimal support, whose support includes the species $Y$.

\eeT

\prf
 Let $X_j$ denote the ACR species and $x_j^*$ denote the ACR value. Since the system has a positive conservation law, we have
\begin{equation}
\label{cons}
\sum_{i=1}^n c_i x_i = x_{tot}\in \mR_+,
\end{equation}
where $c_i > 0$ for all $i = 1, \ldots, n$.  At the positive steady state we can rearrange this to
\[0 < \mathop{\sum_{i=1}^n}_{i\not=j} c_i x_i = x_{tot} - c_j x_j^* < 0\]
which is a contradiction if  $x_{tot}$ is chosen sufficiently small, \satg\ $x_{tot} - c_j x_j^* < 0$. Consequently, there is no positive steady state for such values of $x_{tot}$.
Since the invariant space is a subset of the intersection of \eqref{cons} with the positive orthant, it is compact and therefore must contain a steady state. It follows that this
steady state must be on the boundary of the positive orthant.

Note that the  problem of global dynamical properties of ACR networks,
  called {\bf dynamic ACR}, is an active field of research \cite{feliu2012,Radh2013,cappelletti2020,joshi23,puente2025}.

\section{Examples of NGM stability analysis for CRNs}\lab{s:ACRex}
\subsection{The (5,6,3,2) IDHKP-IDH system~\cite{ShiFei,joshi2024bifunctional}
 does not \saty\ the NGM conditions, but \sats\ its conclusion: IDHKP.nb}
 \lab{s:ShiFei}
The following (5,6,3,2) network
	 \begin{equation}\label{e.ptm}
	       \begin{tikzcd}
S_1 + C_0 \arrow[r,yshift=+0.5ex,"k_1"] & C_1 \arrow[l,yshift=-0.5ex,"k_2"] \arrow[r,yshift=+0.5ex,"k_3"]& S_2 + C_0 & \\
[-0.1in]
S_2 + C_1 \arrow[r,yshift=+0.5ex,"k_4"] & C_2 \arrow[l,yshift=-0.5ex,"k_{5}"] \arrow[r,yshift=+0.5ex,"k_{6}"]& S_1 + C_1 &
      \end{tikzcd}
%\begin{array}{rl}  S_1+ C_0\xrightleftharpoons[k_2]{k_1}
%C_1 \xrightarrow{k_3} S_2+C_0\\ S_2+C_1 \xrightleftharpoons[k_5]{k_4}
%C_2 \xrightarrow{k_6} S_1+C_1 \end{array}
\end{equation}
was introduced in Ref. \cite{ShiFei}, as an example of network satisfying the sufficient
ACR conditions of having deficiency one, and two non-terminal complexes
($C_1, S_2+C_1$), (the terminal complexes above are $(S_2+C_0, S_1+C_1$),
which
  differ in a single species).

  \begin{tcolorbox}[colback=gray!10,title=Software]
The computations below, and for all examples in this paper, are available in symbolic form in the Epid-CRN Mathematica package (GitHub: \url{https://github.com/florinav/EpidCRNmodels}), where scripts  for computing Jacobians, conservation laws, bifurcation conditions, and Routh-Hurwitz tests are provided.
\end{tcolorbox}

	Denote the state vector by $X=[C_0,S_1,C_1,S_2,C_2]^T$. The corresponding ODE can be written as:
	 \begin{gather} \nonumber\dot X=\frac{d}{dt}\bep C_0\\S_1\\C_1\\S_2\\ C_2 \eep=
\left(
\begin{array}{cccccc}
 -1 & 1 & 1 & 0 & 0 & 0 \\
 -1 & 1 & 0 & 0 & 0 & 1 \\
 1 & -1 & -1 & -1 & 1 & 1 \\
 0 & 0 & 1 & -1 & 1 & 0 \\
 0 & 0 & 0 & 1 & -1 & -1 \\
\end{array}
\right)
\bep  k_1 C_0S_1\\C_1 k_2\\C_1 k_3\\C_1 k_4 S_2\\C_2 k_5\\C_2 k_6\eep =
 \\ \bep C_1 \left(k_2+k_3\right)- k_1 C_0S_1\\C_2 k_6-k_1 C_0S_1+C_1 k_2\\C_2 \left(k_5+k_6\right)+ k_1 C_0S-C_1 \left(k_4 S_2+k_2+k_3\right)\\C_2 k_5+C_1 \left(k_3-k_4 S_2\right)\\
C_1 k_4 S_2-C_2 \left(k_5+k_6\right)\eep:=RHS
\end{gather}

\BEN \im The two conservation laws (found for example with EpidCRN)
\be{RAPc}\left(
\begin{array}{c}
 C_0+C_1 +C_2\\
 2 C_2+C_1+S_1+S_2 \\
\end{array}
\right)=\left(
\begin{array}{c} c_{tot}\\
s_{tot}\end{array}
\right)\ee
 imply that the network  admits
a conservation with all coefficients positive, and therefore trajectories remain in a compact, forward-invariant set \cite{feinberg1987chemical} (see lemma 2.1 and discussion).

 \im  Solve[RHS==0,x] finds quickly three  two dimensional equilibria manifolds, one of which is  the rational parametrization:
    \be{RAPi} \bc S_1= \frac{k_2+k_3}{ k_1}\frac{C_1}{C_0}\\
    S_2=  \frac{k_3 \left(k_5+k_6\right)}{k_4 k_6}:= \fr 1 {\mR}
   \\ C_2= \frac{k_3}{k_6} C_1.\ec \ee
Note the second component identifies an ACR species (in agreement with the \cite{ShiFei} criterion), with ACR value $s_2^*=  \frac{k_3 \left(k_5+k_6\right)}{k_4 k_6}$, which was denoted also by $\fr 1 {\mR}$, to emphasize the analogy with the ACR value for the endemic susceptibles, in ME models. The equation $det(M)=0$ confirms the ACR value $s_2^*$.

Solve also finds two boundary \fp\ possibilities: \BEN \im  $\left\{C_0=C_1=C_2= 0\right\},$ which is incompatible with the conservation $C_0+C_1 +C_2= c_{tot}$, and
\im $\left\{S_1=C_1=C_2= 0\right\}$, which yields the  ``DFE"   fixed point
 \be{DFEs}S_1=0,S_2=s_{tot}, C_1=0,C_2=0, C_0=c_{tot} .\ee
 \EEN

    Therefore, the resident species are $\{C_0, S_2\}$ and the invading species are $\{S_1, C_1, C_2\}$.

\im We consider first  the NGM approach, which starts by splitting the RHS of the differential equations for the ‌invading  species $S_1,C_1,C_2$ as follows
\begin{gather*}
\bep -C_0 k_1 S_1+C_1 k_2+C_2 k_6\\C_0 k_1 S_1-C_1 \left(k_4 S_2+k_2+k_3\right)+C_2 \left(k_5+k_6\right)\\C_1 k_4 S_2-C_2 \left(k_5+k_6\right)\eep=\mF-\mV\\:=
\bep 0\\
C_0 k_1 S_1
\\C_1 k_4 S_2\eep-
\bep C_0 k_1 S_1-C_1 k_2-C_2 k_6\\
C_1 \left(k_4 S_2+k_2+k_3\right)-C_2 \left(k_5+k_6\right)
\\C_2 \left(k_5+k_6\right)\eep,
\end{gather*}
and we included in the   ``new infections'' part $\mF$  all nonlinear, positive interactions.  \How we may not apply NGM,
since the condition (A4) of the NGM theorem \cite{Van08},
 which requires $\mV$ to have a  nonnegative  sum for all nonnegative x and y, is not \satd\ here (the sum is $C_0 k_1 S_1+ k_4 C_1  S_2\blue{-C_1 k_3}-C_2 k_6$).

Despite the fact that  the condition (A4) of the NGM theorem \cite{Van08} isn't satisfied, we may use the ``NGM heuristic" from section \ref{s:heu} (simply ignoring the condition (A4) of \cite{Van08}), and  this will turn out to get  the correct  stability domain of the DFE:
 \BEN \im The Jacobian of the  invasion  equations \wrt\ the  invasion  variables (considered without rigorous justification) may be decomposed as
 \begin{gather*} \lab{Jx} J_x=\left(
\begin{array}{ccccc}
 -k_1 C_0& k_2 & k_6 \\
 k_1 C_0& -k_4 S_2-k_2-k_3 & k_5+k_6 \\
 0 & k_4 S_2 & -k_5-k_6 \\
\end{array}
\right)=F-V, \\F=\left(
\begin{array}{ccccc}
 0 & 0 & 0 \\
 k_1 C_0& 0 & 0 \\
 0 & {k_4 S_2} & 0 \\
\end{array}
\right),-V=\left(
\begin{array}{ccccccc}
 -k_1 C_0& k_2 & k_6 \\
 0& -k_4 S_2-k_2-k_3 & k_5+k_6 \\
 0 & 0 & -k_5-k_6
\end{array}
\right)
\end{gather*}
$V$ is diagonally dominant, and so  this is a \regS.

The resulting  \NGM\ has one zero \eig,  and  $R_0$ is the root of  a quadratic equation. The stability domain  $R_0<1$ may be checked  by CAS to be correct.

 Alternatively,  removing the $0$ root and substituting  $u\to x+1$ we arrive to a classic second order \RH\ stability problem, where only the free \coe\ $k_3 \left(k_5+k_6\right)-k_4 k_6 s_{tot}$ may change sign, confirming
 that stability of the DFE occurs when: \be{RH2}\frac{k_3 (k_5+k_6)}{k_4 k_6} > s_{tot} \Eq R_0:= s_{tot} \mR <1, \mR:=\frac{k_4 k_6}{k_3 (k_5+k_6)}.\ee

\EEN

\im  We show next  that classic stability analysis produces the same result. The \ch\ of the full Jacobian
 factors out $u^2$. The resulting third order polynomial (note that we have reduced
 the dimension to 3 without using the  NGM theorem)  further simplifies at the DFE to a polynomial for which  only the  first coefficient may change sign, and this happens at $s_2^*=s_{tot}$. Our CAS confirms that the DFE is LAS when $R_0<1$.

 \im
 Finally, \RH\ analysis  shows that the non-boundary steady state is stable whenever it exists.
\EEN

Due to the conservation constraints, the expression of $C_1$ at the EE is given by 
$$C_1=\frac{C_0 k_1 \left(k_4 k_6 s_{tot}-k_3 (k_5+k_6)\right)}{C_0 k_1 k_4 \left(2 k_3+k_6\right)+\left(k_2+k_3\right) k_4 k_6},$$ which is positive iff 
\be{EEp} s_{tot}\geq \frac{k_3(k_5+k_6)}{k_4 k_6},\ee  precisely the complement of \eqr{RH2}.

%see last cell in IDHKP.nb

 We conjecture
that  the EE  is globally asymptotically  stable
  when  \eqr{EEp} holds.

%\subsection{Computing $R_0$  for several representations of  the Escherichia coli EnvZ-OmpR osmolarity regulator}

\subsection{Examples of  Escherichia coli EnvZ-OmpR models, with ACR species Yp, where  \eqr{R0fr} holds}\lab{s:EnvZs}
In this section,   we denote EnvZ by X, and OmpR by Y.  Note  that
alternations of capital and small letters following each other, like in XdYp, will denote here a single variable, while the same juxtaposition, but using sub-indices, will denote a product $X_d Y_p=X_d *Y_p$.

\subsubsection{A  (5,4,3,1) subnetwork of   EnvZ-OmpR : EnvZ.nb}\lab{s:EnvZ}
\figu{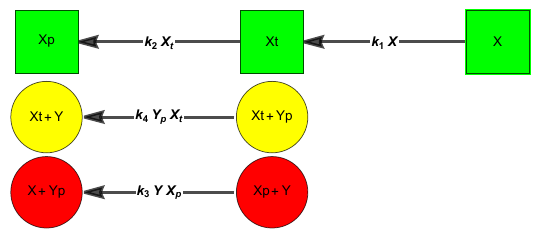}{The FHJ graph of an EnvZ-OmpR network with $5$ species, $4$ reactions and  deficiency $\de=7-3-3=1$.}{1}

This section considers a simplified EnvZ-OmpR  system represented in \ref{f:EnvZgr}
%\begin{equation}
%\label{envzompr}
%\begin{tikzcd}
%X  \arrow[r,"k_1"] & X_t  \arrow[r,"k_2"] & X_p \\[-0.1in]
%X_p + Y \arrow[r,"k_3"] & X  + Y_p \\[-0.1in]
%X_t  + Y_p \arrow[r,"k_4"] & X_t  + Y
%\end{tikzcd}
%\end{equation}
with \sm:
\bea \left(
\begin{array}{cccc}
 -1 & 0 & 1 & 0 \\
 1 & -1 & 0 & 0 \\
 0 & 1 & -1 & 0 \\
 0 & 0 & -1 & 1 \\
 0 & 0 & 1 & -1 \\
\end{array}
\right),\eea rank $3$,  \defi\ $1=7-3-3$.

 The corresponding system of ordinary differential equations is:
\begin{equation}
\label{de}
\frac{d}{dt} \left( \begin{array}{c} X \\ X_t \\ X_p \\ Y \\ Y_p \end{array} \right) = \left( \begin{array}{c} -k_1 X  + k_3 X_p Y \\ k_1 X  - k_2 X_t \\ k_2 X_t  - k_3 X_p Y \\ -k_3 X_p Y + k_4 X_t  Y_p \\ k_3 X_p Y - k_4 X_t  Y_p \end{array} \right)
\end{equation}
%\begin{equation}
%\label{de}
%\left\{ \begin{aligned}
%X' & = -k_1 X  + k_3 X_p Y \\
%X_t ' & = k_1 X  - k_2 X_t \\
%X_p' & = k_2 X_t  - k_3 X_p Y \\
%Y' & = -k_3 X_p Y + k_4 X_t  Y_p \\
%Y_p' & = k_3 X_p Y - k_4 X_t  Y_p
%\end{aligned} \right.
%\end{equation}
(to simplify the notation, we  have abbreviated, as customary,  EnvZ by X, and OmpR by Y). This example  satisfies the sufficient \cite{ShiFei}
ACR conditions,  since it  has deficiency one, and two non-terminal complexes
$X_t, X_t+Y_p$ (see \ref{f:EnvZgr}),
which
  differ in the single species $Y_p$ (which must therefore be  the ACR species).

 %\figu{EnvZOc}{Flow chart corresponding to the system \eqr{de}. }{1}

\BEN \im Our first step is the determination of the flux cone,  generated here by  $\left(
\begin{array}{c}
 1 \\
 1 \\
 1 \\
 1 \\
\end{array}
\right)$, and   the conservation laws $x_{tot} = X  + X_t  + X_p, y_{tot} = Y + Y_p$, (total EnvZ and total OmpR, \resp). Both are found using our script ``cons'', which simply solves the linear kernel equations under non-negativity constraints on the variables.

 \im The second step is trying to get information on the non-negative fixed points. In simple examples like the ones discussed in this paper,
 \lstinline[language=Mathematica]{Solve[RHS==0,var]} (the imperfect, quicker version of Reduce) finds quickly  the equilibria manifolds. In this example, it finds  two boundary ones,  with two common invading components $ X =X_ t = 0$. Under this condition, the steady points must also \saty\  $ X_p Y =0$. The first possibility $ X_p  =0$ is incompatible with the positivity
of the first conservation, and the second,  $ Y= 0$, yields the DFE $$X  =  X_t  =0=Y,  X_p  =x_{tot},  Y_p =y_{tot},$$
with resident species set $\{ X_p, Y_p\}$.

\beR When Solve doesn't finish in reasonable time, it can be replaced by a search for boundary points, which are in fact typically easier to find (even when interior points aren't). %as is the custom
\eeR

 Solve identifies here also a third fixed point with  all \coords\ positive, parametrized by $x  ^*,x_p  ^*$,  \satg:
\[x_t  ^*=\frac{k_1 x  ^*}{k_2}, y  ^*= \frac{k_1 x  ^*}{k_3 x_p  ^*}, y_p ^*=\frac{k_2}{k_4} \leq y_{tot},\]
the last component $y_p ^*$ being the ACR species.
Note the last natural inequality, which turns out here to imply  both uniqueness of the interior point within each invariant set, parametrized  by $x_{tot}, y_{tot}$,  (so, finally, this is an example of trivial ACR), and stability, as shown next.

\im The next  step is trying to apply the NGM theorem (and even when it does not apply,  we get useful information by applying it formally -- examples below for further remarks).

Here, the set of all infectious/invading species is $\{ X, X_t , Y \}$ (see first step), and  the RHS of the
infection equations is
\[RHS_i=\bep k_3 Y X_p-k_1 X\\k_1 X-k_2 X_t\\k_4 Y_p X_t-k_3 Y X_p\eep=\mF-\mV:=\bep
 k_3 Y X_p\\0\\k_4 Y_p X_t\eep-\bep k_1 X\\k_2 X_t-k_1 X\\k_3 Y X_p\eep\]
 Here, $RHS_i$ has been split into the part of ``new infections" $\mF$
containing  all the interactions with positive sign \cite{AABJ}, and  the rest, denoted by $-\mV$, with the purpose of replacing the study of the spectrum of Jacobian $J_i$ \eqr{jI}
by that of the ``next generation matrix" \eqr{K}.

 Since the sum of the rows of $\mV$,  $k_2 X_t+k_3 Y X_p$ is unconditionally positive, and the other conditions of \cite{Van,Van08} also apply, we may apply the NGM theorem.

The derivative of $RHS_i$ \wrt\ the infection variables is:
\begin{gather}\label{jI} \nonumber J_i=\left(
\begin{array}{ccc}
 -k_1 & 0 & k_3 X_p   \\
 k_1 & -k_2 & 0 \\
 0 & k_4 Y_p   & -k_3 X_p   \\
\end{array}
\right)=F-V=\left(
\begin{array}{ccc}
 0 & 0 & k_3 X_p   \\
 0 & 0 & 0 \\
 0 & k_4 Y_p   & 0 \\
\end{array}
\right)-\\
\left(
\begin{array}{ccc}
 k_1 & 0 & 0 \\
 -k_1 & k_2 & 0 \\
 0 & 0 &
 k_3 X_p   \\
\end{array}
\right) \Lra
\end{gather}

\be{K} K = FV^{-1} = \left( \begin{array}{ccc}
0 & 0 & 1 \\
0 & 0 & 0 \\
\displaystyle{\frac{k_4 y_{tot}}{k_2}} & \displaystyle{\frac{k_4 y_{tot}}{k_2}} & 0
\end{array}
\right)
\ee
where the resident species $X_p, Y_p$ are evaluated
at the ``disease-free state" (with $X  = X_t = Y = 0$),
i.e. $X_p = x_{tot}, Y_p = y_{tot}$ (this is implemented in our script NGM).
\im The next step consists in attempting to
factor  the \ch\  of $K$, and in removing ``stable factors" (i.e. factors which may not have \eig s with positive real part). It seems, at least for ME models and for the ACR models \satg\ the \cite{ShiFei} conditions discussed in this chapter,   that the \ch\ of $K$ factors more often than that of $J_i$.

 In our current example, the 3rd degree \ch\ of $K$ is $ch(u)=u (k_2 u^2-k_4 Y_p)$. After removing the factor $u$,    the remaining second degree polynomial has symmetric roots, the positive one being
$$R_0=\sqrt{\frac{k_4 y_{tot}}{k_2}}=\sqrt{\frac{ y_{tot}}{y_p  ^*}}.$$

Finally, the condition  $R_0 > 1$ under which the stability of the boundary  disease-free steady state is lost is precisely
\be{ACRi}y_{tot} > \frac{k_2}{k_4}=y_p^*.\ee

\beR When the  \ch\ has higher degree, we may also apply,  alternatively, the RH criteria to the shifted polynomial $p(x)=ch(x+1)$, like in the previous section. In our current example,
the shifted second degree polynomial is $k_2 x^2+2 k_2 x+k_2-k_4 Y_p$, and RH recovers
the condition $k_2-k_4 Y_p\ge 0$. The computational advantage of this approach is that the single condition $R_0 \leq 1,$  with $R_0$ which may be the root of a high order polynomial, may be replaced by several RH conditions. These may be simpler sometimes, as will be illustrated in the   examples below.
\eeR

\EEN

\beR Here, $R_0$ may be  obtained  also using the Jacobian $J_i$ (whose \ch\ does not factor).
Indeed, the non-zero roots $\l$ of this \ch\ turn out to \saty\
$$\left(\lambda/k_1 +1\right) \left(\lambda/k_2+1\right) \left(\lambda/(k_3 X_p  )+1 \right)=  k_4  Y_p  /k_2,$$
and an eye inspection (or \Ma) checks that a crossing of $0$ by $\l$ occurs  iff $R_0=1$; however, the computation which uses the NGM matrix $K$ is easier, possibly due to the larger number of $0$'s, as compared to using the reduced Jacobian $J_i$, which does not factor.
\eeR

\iffalse 
\beR \lab{r:Env}  The stability of the endemic point (when it exists) may also be
quickly checked, via:
\begin{lstlisting}
Reduce[Append[Join[ct, Thread[co > 0]], H3=co_2 co_3-co_1 co_4 > 0]]
\end{lstlisting}
where  ct denote all the positivity conditions (parameters and variables),
co denote the coefficients of the Jacobian reduced to three dimensions by dividing by $\l^2$, and H3 is the third  Hurwitz determinant.
\fi
  In conclusion,
we have shown that for the simplified EnvZ-OmpR that a typical scenario for ME models \cite{Diek,Van,Van08} holds: for $R_0 <1 \Eq y_{tot} < y_p^*$ the DFE is locally stable, and for $R_0 \ge 1 \Eq y_{tot} \ge y_p^*, $ the second endemic point enters the positivity domain and takes over the ``stability relay" from the DFE.

%\subsubsection{A (8,9,,) version of  EnvZ-OmpR studied by \cite{AliciaMessi}}
%The CRN
%${\bc X <-> X_t-> X_p,\\
    %  X_p+Y<-> XpY -> X+Y_p,\\
     % X_t+Y_p<-> XtYp -> X_t+Y\ec}$

\subsubsection{A (8,11,6,1) subnetwork of  EnvZ-OmpR \cite{cappelletti2020hidden}: EnvZCap.nb}\lab{s:EnvZCap}

 We present  now  a subnetwork of the  EnvZ-OmpR model, which was studied in Ref. \cite{cappelletti2020hidden}, and which  is obtained by ignoring the last line in the more
 general EnvZ-OmpR model  studied in Refs. \cite{ShiFei,karp2012complex}. %and in  also \eqr{EJ} below:
%\be{Cap}
%      \begin{tikzcd}
%X_d \arrow[r,yshift=+0.5ex,"k_1"] & X \arrow[l,yshift=-0.5ex,"k_2"] \arrow[r,yshift=+0.5ex,"k_3"]& X_t \arrow[l,yshift=-0.5ex,"k_4"] \arrow[r,"k_5"]& X_p\\[-0.1in]
%X_p + Y \arrow[r,yshift=+0.5ex,"k_6"] & XpY \arrow[l,yshift=-0.5ex,"k_7"] \arrow[r,yshift=+0.5ex,"k_8"]& X + Y_p & \\
%[-0.1in]
%X_d + Y_p \arrow[r,yshift=+0.5ex,"k_9"] & XdYp \arrow[l,yshift=-0.5ex,"k_{10}"] \arrow[r,yshift=+0.5ex,"k_{11}"]& X_d + Y &
%      \end{tikzcd}
%%\schemestart
%%EnvZ-D  \arrow(1--2){<=>[*0{$k_1$}][*0{$k_2 $}]} EnvZ \arrow(@2--3){<=>[*0{$k_3 $}][*0{$k_4$}]} EnvZ-T \arrow(@3--4){->[*0{$k_5$}]} EnvZ-P
%%\arrow(@1.south east--.north east){0}[-90,.50]
%%EnvZ-P \+ OmpR \arrow(5--6){<=>[*0{$k_6$}][*0{$k_7$}]} EnvZ-OmpR-P \arrow(@6--7){->[*0{$k_8$}]} EnvZ \+ OmpR-P
%%\arrow(@5.south east--.north east){0}[-90,.50]
%%EnvZ-D \+ OmpR-P \arrow(8--9){<=>[*0{$k_9$}][*0{$k_{10}$}]} EnvZ-OmpR-D-P \arrow(@9--10){->[*0{$k_{11}$}]} EnvZ-D \+ OmpR
% %\schemestop
%\ee

%EnvZCap.nb
\figu{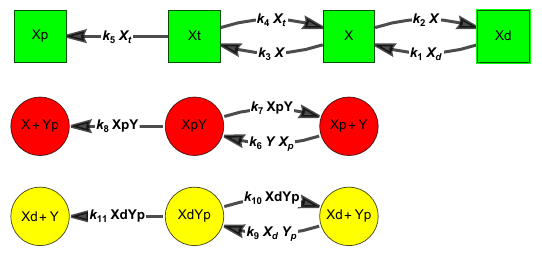}{The FHJ graph of an (8,11,6,1)  EnvZ-OmpR model with $8$ species, $11$ reactions and  deficiency $\de=10-3-6=1$.}{1}

\beR  The conditions from Ref \cite{ShiFei} identify OmpR$=Y_p$ as an ACR species.
\eeR

Ordering  the species (and the complexes) according to their appearance order yields the ODE: \beq &&\lab{ODEC}\frac{d}{dt} \bep X_d\\
 X\\
  X_t\\
   X_p\\
   Y\\
 XpY\\
  Y_p\\
   XdYp\eep=
   \bep-X_d \left(k_9 Y_p+k_1\right)+k_2 X+
   \left(k_{10}+k_{11}\right) {XdYp}
   \\k_1 X_d+k_4 X_t-X \left(k_2+k_3\right) +k_8 {XpY}\\k_3 X-\left(k_4+k_5\right) X_t\\-k_6 Y X_p+k_5 X_t+k_7 {XpY}\\-k_6 Y X_p+k_{11} {XdYp}+k_7 {XpY}\\k_6 Y X_p-\left(k_7+k_8\right) {XpY}\\-k_9 X_d Y_p+k_{10} {XdYp}+k_8 {XpY}
   \\k_9 X_d Y_p-\left(k_{10}+k_{11}\right) {XdYp}\eep\\&&=\left(
\begin{array}{c}
 0 \\
 0 \\
 0 \\
 0 \\
 0 \\
 k_6 Y X_p \\
 0 \\
 k_9 X_d Y_p \\
\end{array}
\right)-
\left(
\begin{array}{c}
 k_9 X_d Y_p+k_1 X_d-k_2 X -\left(k_{10}+k_{11}\right) {XdYp}\\
 -k_1 X_d-k_4 X_t+(k_2+k_3) X -k_8 {XpY}\\
 \left(k_4+k_5\right) X_t-k_3 X \\
 k_6 Y X_p-k_5 X_t-k_7 {XpY} \\
 k_6 Y X_p-k_{11} {XdYp}-k_7 {XpY} \\
 (k_8 +k_7) {XpY} \\
 k_9 X_d Y_p-k_{10} {XdYp}-k_8 {XpY} \\
  \left(k_{10}+k_{11}\right){XdYp} \\
\end{array}
\right).\eeq

%\figu{EnvZCapc}{Flow chart corresponding to the ODE \eqr{ODEC} where the sum of the RHS yields $-k_9 X_d Y_p-k_6 Y X_p+\left(k_{10}+k_{11}\right) XdYp+\left(k_7+k_8\right) XpY$, and the node $0$ corresponds to the \blue{exterior outputs and inputs. } }{1}

The system \eqr{ODEC} admits the following DFE:
$$ \pr{X=0,X_d=0,X_t=0,Y=0,XpY=0,XdYp=0}. $$

After removing  the fourth and seventh row in $\mV$, which correspond to resident variables, we find that the sum of the rows in
$$\mV=\left(
\begin{array}{c}
 k_9 X_d Y_p+k_1 X_d-k_2 X -\left(k_{10}+k_{11}\right) {XdYp}\\
 -k_1 X_d-k_4 X_t+(k_2+k_3) X -k_8 {XpY}\\
 \left(k_4+k_5\right) X_t-k_3 X \\

 k_6 Y X_p-k_{11} {XdYp}-k_7 {XpY} \\
 (k_8 +k_7) {XpY} \\

  \left(k_{10}+k_{11}\right){XdYp} \\
\end{array}
\right),$$ given by  $k_9 X_d Y_p
 +k_5 X_t+
k_6 Y X_p
 $ is unconditionally positive, and  we may apply the NGM theorem.

\BEN \im
It can be checked that $rank(\Gamma)=6$ and the deficiency is
$\delta=m-rank(\Gamma)-\ell=10-6-3=1.$
The CRN is conservative, and the conservations are generated by:
\begin{equation}\label{eq:cons_envz}
\left\{X_d+X_p+X_t+X+{XdYp}+{XpY},Y_p+{XdYp}+{XpY}+Y\right\}
\end{equation}
(corresponding to the fact that the total mass of the chemical species containing some form of EnvZ is conserved, as well as the total mass of the chemical species containing some form of OmpR). The cone of positive fluxes has dimension 4.

 \im \Ma\ yields the following rational parametrization (RAP) parametrized by $X_d,X_p$, which is computed   using Solve[RHS==0] without specifying the ``var":
\be{SCap}\bc X= \frac{k_1 X_d}{k_2},X_t= \frac{k_1 k_3 X_d}{k_2 \left(k_4+k_5\right)},\\Y= \frac{k_1 k_3 k_5 \left(k_7+k_8\right) X_d}{k_2 \left(k_4+k_5\right) k_6 k_8 X_p},\text{XpY}= \frac{k_1 k_3 k_5 X_d}{k_2 \left(k_4+k_5\right) k_8},\\Y_p^*= \frac{k_1 \left(k_{10}+k_{11}\right) k_3 k_5}{k_{11} k_2 \left(k_4+k_5\right) k_9},\text{XdYp}= \frac{k_1 k_3 k_5 X_d}{k_{11} k_2 \left(k_4+k_5\right)}\ec\ee

\im  The NGM approach, applied to the 6D projection on the DFE coordinates, yields 4 eigenvalues equal to 0, and two satisfying a quadratic equation (again, the \ch\ of $K$ factors, while that of $J_i$ doesn't).
After shifting the polynomial, as explained in point 4. of the previous section,
we obtain a polynomial whose free coefficient at the DFE is
$$co_1=k_8 \left(k_1 \left(k_{10}+k_{11}\right) k_3 k_5-k_{11} k_2 \left(k_4+k_5\right) k_9 y_{tot}\right),$$
and the other two coefficients are positive.

 In conclusion, the stability domain of the DFE, obtained from
  $co_1>0$, is:
\be{ECint} y_{tot} \leq \frac{k_1 \left(k_{10}+k_{11}\right) k_3 k_5}{k_{11} k_2 \left(k_4+k_5\right) k_9} \Eq R_0:=\fr{y_{tot}}{Y_P^*}\leq 1. \ee

\EEN

\section{Network translation and WR-ZD realizations: the  (4,5,3,2) mass action system with ACR   \cite[(1)]{tonello2018network}: Tonello.nb}\lab{s:Ton}

In this section, we mention one idea which might turn out useful for the non-reversible models of ME. There exists one idea
of CRNT, the embedding of the \HJF\ into Euclidean space, which has not been  mentioned until now.  Neither  was the question of how moving the graph vertices around (and maybe allowing them to collide) might affect the results. This  question was  addressed in the theory of network translation, initiated by Johnston  \cite{John,tonello2018network,JMP,hernandez2022framework}.

 Network translation of a system with MAK to a WR-ZD with GMAK (generalized mass action kinetics) helps understanding, among other things,  rational parametrizations related to the  matrix tree theorem (see for example
 Ref. \cite{CoxSl}) which appear in certain MAKs. Consider the MAK \cite[(1)]{tonello2018network}
 \begin{equation}\label{eq:proper_acr_intro}
      \begin{tikzcd}[row sep=small, column sep=small]
          & A + C & C \arrow[rd, "r_3"] & & \\
         A + B \arrow[ru, "r_1"] \arrow[rd, "r_2"'] & & & A \arrow[r, "r_5"] & B. \\
        & A + D & D \arrow[ru, "r_4"'] & &
      \end{tikzcd}
\end{equation}

Ordering  the species lexicographically %(and the complexes) according to their appearance order
yields the ODE: \beq &&\lab{ODET}\frac{d}{dt} \bep A\\
 B\\
 C\\
   D\eep=
   \bep-k_5 A+k_3 C+ k_4 D\\-A \left(k_1 B+k_2 B-k_5\right)\\
   k_1A B - k_3 C\\k_2A B - k_4 D\eep\eeq
   with \sm\ $\G=\left(
\begin{array}{ccccc}
 0 & 0 & 1 & 1 & -1 \\
 -1 & -1 & 0 & 0 & 1 \\
 1 & 0 & -1 & 0 & 0 \\
 0 & 1 & 0 & -1 & 0 \\
\end{array}
\right)$.

\BEN \im
It can be checked that $rank(\Gamma)=3$, and the deficiency is
$\delta=n_C-rank(\Gamma)-\ell=7-3-2=2$ (note this is our first example of deficiency $2$).
 The CRN has one conservation
$
A+B+C+D
$, and  the cone of positive fluxes has dimension 2.

 \im \Ma\ finds two fixed points
\be{Ton}\bc A= 0,B=n_{tot},C= 0,D= 0\\ B^*= \frac{k_5}{k_1+k_2},C= \frac{ k_1 k_5}{\left(k_1+k_2\right) k_3}A,D= \frac{ k_2 k_5}{\left(k_1+k_2\right) k_4}A\ec,\ee
which suggests that the \MTT\ might be at work.
The second solution under the extra constraint $A+B+C+D=n_{tot}$,
\[\bc A=\left(\left(k_1+k_2\right) n_{tot}-k_5\right)\frac{k_3 k_4 }{k_1 k_4 \left(k_3+k_5\right)+k_2 k_3 \left(k_4+k_5\right)},B^*= \frac{k_5}{k_1+k_2}\\C=\left(\left(k_1+k_2\right) n_{tot}-k_5\right)\frac{k_1 k_4 k_5 }{\left(k_1+k_2\right) \left(k_1 k_4 \left(k_3+k_5\right)+k_2 k_3 \left(k_4+k_5\right)\right)}\\D=\left(\left(k_1+k_2\right) n_{tot}-k_5\right)\frac{k_2 k_3 k_5 }{\left(k_1+k_2\right) \left(k_1 k_4 \left(k_3+k_5\right)+k_2 k_3 \left(k_4+k_5\right)\right)}\ec\] is
positive iff $n_{tot} >B^*$.
 %For the NGM method, we consider the ``rank 1'' matrices

  Let's consider the following decomposition of the infectious equations in \eqr{ODET} (produced by our NGM script from Epid-CRN):
  \iffalse
\[\blue{\mF-\mV = \bep k_3 C +k_4 D\\0\\0\eep -\bep k_5 A\\k_3 C-k_1 A B \\ k_4 D- k_2 A B \eep \Rightarrow } F=\left(
\begin{array}{ccccc}
 0 & k_3 & k_4 \\
 0 & 0 & 0 \\
 0 & 0 & 0 \\
\end{array}
\right),  V=\left(
\begin{array}{ccccccc}
 k_5 & 0 & 0 \\
 -k_1 B & k_3 & 0 \\
 -k_2 B & 0 & k_4
\end{array}
\right),\]
\blue{from the previous decomposition note that $\sum \mV_i=k_5 A + k_3 C+k_4 D- A B(k_1+k_2)$ which doesn't satisfy the condition \cite[(A4)]{Van08}, so the NGM cannot be applied.}
\fi
{
\[\mF-\mV=\bep 0\\ k_1 A B\\ k_2 A B\eep -\bep k_5 A-k_3 C -k_4 D\\ k_3 C\\ k_4 D\eep\]
The condition \cite[(A4)]{Van08} is satisfied  since $\sum \mV=k_5 A >0$ (and all the other conditions are \satd\ as well). Now
\[F=\left(
\begin{array}{ccc}
 0 & 0 & 0 \\
 B k_1 & 0 & 0 \\
 B k_2 & 0 & 0 \\
\end{array}
\right), V= \left(
\begin{array}{ccc}
 k_5 & -k_3 & -k_4 \\
 0 & k_3 & 0 \\
 0 & 0 & k_4 \\
\end{array}
\right)\Rightarrow K=\left(
\begin{array}{ccc}
 0 & 0 & 0 \\
 \frac{B k_1}{k_5} & \frac{B k_1}{k_5} & \frac{B k_1}{k_5} \\
 \frac{B k_2}{k_5} & \frac{B k_2}{k_5} & \frac{B k_2}{k_5} \\
\end{array}
\right).\]
The cubic \ch\ of $K$ is $\mbox{ch}(u)=\frac{u^2 \left(n_{tot} (k_1+ k_2)-k_5 u\right)}{k_5}$. After
removing the factor $u^2$, the remaining first order polynomial has one positive root $\fr{n_{tot}}{B^*}$
implying that the DFE is stable iff $R_0:=\fr{n_{tot}}{B^*} \leq 1$. }
%This yields $R_0=\fr{n_{tot}}{B^*}$,
Also, the Jacobian at the EE reveals that it is always stable, when  $R_0>1$.
\EEN

To better understand this model, the authors propose studying in parallel the GMAK
\begin{equation}\label{eq:proper_acr_intro_translated}
      \begin{tikzcd}[row sep=small, column sep=small]
        & & \mbox{\ovalbox{$\begin{array}{c} A+C \\ (C) \end{array}$}} \arrow[rd, "\tilde{r}_3"] & & & \\
& \mbox{\ovalbox{$\begin{array}{c} A+B \\ (A+B) \end{array}$}} \arrow[ru, "\tilde{r}_1"] \arrow[rd, "\tilde{r}_2"'] & & \mbox{\ovalbox{$\begin{array}{c} 2A \\ (A) \end{array}$}}. \arrow[ll, "\tilde{r}_5"] \\
        & & \mbox{\ovalbox{$\begin{array}{c} A+D \\ (D) \end{array}$}} \arrow[ru, "\tilde{r}_4"'] & & &
      \end{tikzcd}
    \end{equation}
    In \eqref{eq:proper_acr_intro_translated}, the lower (kinetic) complexes, displayed in parenthesis,  correspond to the original source complexes \cite{John}, and the upper (stoichiometric) complexes correspond to the source complexes adjusted by  ``translation complexes" (in this case,   $A$ was added to $r_3$, $r_4$, and $r_5$), with the net result of gluing some of the product complexes to source ones, and create this way a WR system with GMAK .

    Note the translated CRN has ZD, and its \fp\ formulas may be interpreted via the \MTT, applied to the graph \eqref{eq:proper_acr_intro_translated} \cite[Thm 2-3]{tonello2018network}.

\section{Conclusions}
This essay is a continuation of recent efforts of ourselves and others to investigate relations between Chemical Reaction Networks (CRN) and mathematical Epidemiology(ME).

In this work, we reinterpret the Next Generation Matrix (NGM) method — central in mathematical epidemiology — within a more general structural framework for CRNs, focusing on Jacobian splittings at boundary fixed points. We apply this to several CRNs with absolute concentration robustness (ACR)  and illustrate symbolic bifurcation analysis using our Mathematica package ``\textbf{EpidCRN.m}".

The essay has four themes.
\BEN \im
Firstly, we show that the fundamental NGM stability result for the local stability of the DFE,
which is little known outside ME, turns out useful for   CRN models with  ACR. This class of models, at least in the examples we studied,  is close qualitatively to  ME models, in the  sense that they have an unique boundary equilibrium
and a unique strictly positive one, and the latter enters the positive domain and becomes \LAS\ (LAS) precisely when a condition similar to the
famous $R_0 >1$ is satisfied, where $R_0$ is the basic reproduction number of ME. Thus, for these ``ME type models",  a ``relay phenomena" takes place: precisely when the DFE loses stability, a new fixed point enters the domain,  and  takes over. This phenomena has been known as \STP\ (STP) \cite{Shuai13}, in cases where the  (LAS) holds actually as \GAS\ (GAS).
\im Secondly, we show that the ACR examples reveal new aspects of NGM, which were not previously made clear
  in ME. For example,   the NGM method gets sometimes the right answer, even though the conditions of
the NGM theorem \cite{Van08} are not all satisfied --see Section \ref{s:ShiFei}. Thus, we suggest   a heuristic approach extending the NGM theorem \cite{Van08}, the validity conditions for which are yet to be studied in the future.
\im We suggest several open problems, for example to  provide a definition of ME models as a subset of CRN models, which encompasses the majority of the ME models of practical interest -- see Open Problem \eqr{o:MEm}.
\im Last but not least, we  offer in the associated GitHub site \url{https://github.com/florinav/EpidCRNmodels }  a Mathematica package, Epid-CRN,  which
  is addressed to researchers of both disciplines, and  provide illustrative notebooks.
Our package may also be used to study  easy cases of  analogue continuous time Markov chain (CTMC) ME and CRN  models, as explained
in  the Arxiv version of the paper, called ``\textit{Investigating synergies  between
  chemical reaction networks (CRN)
  and  mathematical epidemiology (ME), using the  Mathematica package Epid-CRN}".
\EEN

The case studies in Section 3 illustrate the four themes above in diverse CRN contexts.

\beR This paper has a  follow-up paper, where we develop further the x–y theory more fully, and offer also some symbolic implementations of our EpidCRN scripts in Julia. \eeR 

\beO It has been observed in Ref. \cite{AABJ} and before  that sometimes selecting only
a subset of the infection/invading species leaves $R_0$ unchanged, while reducing the computations. Understanding when an ``$R_0$ sufficient minimal infection set", different form the total infection set, exists, is an open problem, which might be related to the concept of siphons of the associated Petri net \cite{AdLSpetri}. \eeO

This suggests that a formal theory of minimal infection sets, possibly based on siphon structure or Petri net representations \cite{AdLSpetri}, might enhance the predictive power of the NGM heuristic.

As a final note, we mention that for now public software siphon detection in the context of symbolic positive ODEs seems to   be
 available only in CoNtRol \cite{donnell2014control} and in the LEARN package \cite{malirdwi_LEARN}, and that  further work on that seems useful (for example for rendering accessible the algorithm based on primary decomposition of binomial ideals announced in \cite{shiu2010siphons,CraStu}). Python and
 Julia versions are also highly desirable.

 With that purpose in mind, we include in the next section some  nontrivial
 Mathematica scripts.

\section{Some nontrivial Mathematica Scripts in EpidCRN}\lab{sec:symbolic}
\BEN \im  The first nontrivial script of our package is the extraction of all syntactically positive terms in a symbolic matrix. Expression replacement is a tricky operation, which  depends on the internal representations of each CAS.

 The following test example, and the correct extraction  were furnished to us by the user @Michael E2 in
 https://mathematica.stackexchange.com/questions/312334 and mathematica.stackexchange.com/questions/286500

 $$M=\left(
\begin{array}{ccc}
 -\text{ai}-\text{ar}+\text{ba} s-\mu  & \text{bi} s & e^{-2 x} \\
 \frac{\text{ai}}{\sqrt{2}}-3 & -\text{$\gamma $i}-\delta -\mu  & -y \\
\end{array}
\right)$$

The extraction of positives function is:
\begin{verbatim}
posM= Replace[#,{_?Negative->0,e_:>Replace[Expand[e],
 {Times[_?Negative,_]->0,t_Plus:>Replace[t,_?Negative|Times[_?Negative,_]->0,1]}]},
 {2}]&;
pM=posM[M] (*
{{ba s,        bi s,  E^(-2 x)},
 {ai/Sqrt[2],  0,     0       }}
*)
\end{verbatim}
\im
When this does not yield  a \regS, one may iteratively extract one rank F's with the following script:
\begin{verbatim}
 FposEx=With[{pos=First@SparseArray[#]["NonzeroPositions"]},
 SparseArray[{pos->Extract[#,pos]},Dimensions@#]]&;
 posEx[pM]
\end{verbatim}
until a \regS\ is found.
\im A third useful script  provides the checking that a long expression, too long to be visualized, is syntactically positive:
\begin{verbatim}
 onlyP[m_]:=m//Together//(*put polynomials in standard form*)
NumeratorDenominator//(*get polys*)Map@CoefficientArrays//
(*get coefficients of polys*)
ReplaceAll[sa_SparseArray:>sa["NonzeroValues"]]//
(*get nonzero coeffs*)Flatten//(*preconditioner for AllTrue*)
AllTrue[#,NonNegative]& (*has 0 if constant term is zero*);
\end{verbatim}
\im We can count the number of  syntactical minuses by:
\begin{verbatim}
 countMS[m_]:=m//Together//(*put polynomials in standard form*)
NumeratorDenominator//(*get polys*)Map@CoefficientArrays//
(*get coefficients of polys*)
ReplaceAll[sa_SparseArray:>sa["NonzeroValues"]]//
(*get nonzero coeffs*)Flatten//(*preconditioner for AllTrue*)
Count[#, _?Negative] &;
\end{verbatim}
\im removes inequalities  in a list with RHS=0 and LHS of one pure symbol
\begin{verbatim}
 seZF[so_]:=Module[{res},res=Select[so,!MatchQ[#,h_[lhs_,0]/;
MemberQ[{Greater,GreaterEqual,Less,LessEqual},h]&&MatchQ[lhs,_Symbol]]&];
If[res==={},True,res]]
\end{verbatim}

\im One of the most challenging scripts is the RUR, useful for the analysis of the purely endemic point. Here, we attempt to solve rationally some
(usually $\x$) variables with respect to the rest, and then eliminate them, aiming to obtain polynomial equations for the remaining \var s. One difficulty is that the Solve might produce several rational solutions, some of then containg $0$-s, which should be eliminated first, using seZF. The ideal situation, and the only one implemented for now, is when one can remove just one equation, and solve \wrt\ its associated variable, with index taken as 1 by default.
\begin{verbatim}
 RUR[mod_, ind_:{1}, cn_ : {}] (*ind is a list*):=
Module[{RHS, var, par, elim,ratsub,pol,ln,rat1},
       RHS = mod[[1]]/.cn; var = mod[[2]]; par = mod[[3]];
       elim = Complement[Range[Length[var]], ind];
       ratsub = seFZ[Solve[Delete[Thread[RHS == 0], ind],
       var[[elim]]]];
       pol =Numerator[Together[RHS//.ratsub]];
       rat1=Append[(ratsub/.var[[ind]]->1),var[[ind]]->1];
    {ratsub, pol,rat1}
      ]
\end{verbatim}
\EEN

\section*{Declarations}

\indent \textbf{Funding: }  MDJ is supported by National Science Foundation Grant No. DMS-2213390.

\textbf{Conflict of Interest:} The authors have no competing interests to declare that are relevant to the content of this article.

\textbf{Acknowledgements:}  We thank Murad Banaji,  Andrei  Halanay, Daniel Lichtblau, Janos Toth, Nicola Vassena  and James Watmough for useful
advice.

 \bibliographystyle{amsplain}%{apalike}

\bibliography{references}

\end{document}
## 5. Siphons and Stability Relays
- Add a brief subsection clarifying the relation of critical siphons to boundary fixed points. Suggest:
> “A critical siphon corresponds to a set of species that vanishes at a boundary fixed point and remains at zero across parameter ranges, unless another siphon is triggered. This interpretation is heuristic, and further clarification from the CRN literature is needed.”

florian nill refers to the per susceptible analog of the reproduction number as the replacement number

 In particular, we have a Mathematica package EpidCRN, some  computational procedures are presented in Sec. 6. I tried  to translate them in  Julia and Python, with help from  Chat and Claude, but for now the task  seemed too hard .